\documentclass[onecollarge,natbib]{svjour2}
\bibpunct{[}{]}{;}{n}{}{,} 
\smartqed  
\usepackage{graphicx}
\journalname{Few-Body Systems}

\usepackage{amssymb}
\usepackage[intlimits]{amsmath}
\usepackage{amsfonts}
\usepackage{dsfont}
\usepackage{slashed}
\usepackage{units}
\usepackage{wrapfig}

      \newcommand{\conjg}[1]{\ensuremath{\hspace{1pt}\overline{\hspace{-1pt}#1\hspace{-1pt}}}\hspace{1pt}}

\usepackage[usenames]{color}
\usepackage{colortbl}

\definecolor{webgreen}{rgb}{0,0.75,0}
\definecolor{webred}{rgb}{0.75,0,0}
\definecolor{webblue}{rgb}{0,0,0.75}
\definecolor{darkblue}{rgb}{0,0,0.6}
\definecolor{darkgreen}{rgb}{0,0.5,0.5}
\definecolor{darkpurple}{rgb}{0.5,0,0.5}
\definecolor{darkorange}{rgb}{1,0.5,0}
\definecolor{darkgrey}{rgb}{0.4,0.4,0.4}
\definecolor{lgray}{rgb}{0.95,0.95,0.95}
\definecolor{lgreen}{rgb}{0.95,1.00,0.90}
\definecolor{lred}{rgb}{1.00,0.90,0.80}
\definecolor{lblue}{rgb}{0.2,0.35,1.00}
\definecolor{shadecolor}{rgb}{1.00,0.92,0.82}

\usepackage[colorlinks=true,linkcolor=lblue,citecolor=lblue,urlcolor=lblue]{hyperref}

\def\longlonglongrightarrow{\relbar\joinrel\relbar\joinrel\relbar\joinrel\relbar\joinrel\relbar\joinrel\relbar\joinrel\rightarrow}

\begin{document}

\title{More about the light baryon spectrum
}


\author{Gernot Eichmann     
}


\institute{G. Eichmann \at
              Institut f\"ur Theoretische Physik, Justus-Liebig-Universit\"at Giessen, 35392 Giessen, Germany \\
              \email{gernot.eichmann@theo.physik.uni-giessen.de}           
}

\date{Received: date / Accepted: date}

\maketitle

\begin{abstract}
  We discuss the light baryon spectrum obtained from a recent quark-diquark calculation, implementing non-pointlike diquarks
  that are self-consistently calculated from their Bethe-Salpeter equations.
  We examine the orbital angular momentum content in the baryons' rest frame and highlight the fact that baryons carry all possible values
  of $L$ compatible with their spin, without the restriction $P=(-1)^L$ which is only valid nonrelativistically.
  We furthermore investigate the meaning of complex conjugate eigenvalues of Bethe-Salpeter equations, their possible connection
  with `anomalous' states, and we propose a method to eliminate them from the spectrum.
\keywords{Nucleon resonances \and Bethe-Salpeter equation \and Faddeev equation \and Complex eigenvalues \and Anomalous states}
\end{abstract}

\section{Introduction}
\label{intro}

  The light baryon spectrum has traditionally been a benchmark of our theoretical understanding of Quantum Chromodynamics (QCD),
  and it continues to be an exciting area of research in light of new advances with photo- and electroproduction experiments~\cite{Klempt:2009pi,Tiator:2011pw,Aznauryan:2012ba,Crede:2013sze}.
  In recent years lattice QCD has successfully charted the low-lying hadron spectrum, but
  lattice determinations of baryon resonances are still in their early stages~\cite{Gockeler:2012yj,Wilson:2015dqa,Lang:2016hnn,Wu:2016ixr}. Recent progress has also been made
  using covariant bound-state equations~\cite{Bashir:2012fs,Segovia:2015hra,Eichmann:2016yit}.
  Here the starting point is the covariant three-body Faddeev equation for baryons~\cite{Eichmann:2009qa},
  whose kernel contains all possible quark-gluon interactions that are encoded in QCD's $n$-point functions.
  The latter can be determined by their Dyson-Schwinger equations, which are the quantum equations of motions of QCD and therefore exact.
  Due to the complexity of the problem most studies so far have employed comparatively simple kernels which, in particular, do not implement decay mechanisms:
  the hadrons obtained so far correspond to poles on the real axis in the associated scattering amplitudes.
  On the other hand, the absence of meson-baryon interactions that would shift these poles into the complex plane can also be viewed as an advantage, as it provides
  insight into the `quark core' of a baryon that is stripped of its meson cloud.

  A possibility to simplify the three-body problem is to introduce diquarks.
  There is a wide range of theoretical quark-diquark studies including quantum-mechanical models~\cite{Klempt:2009pi,Anselmino:1992vg}, light-front holographic QCD~\cite{Brodsky:2014yha},
  or quantum-field theoretical approaches within QCD where diquarks appear as intermediate structures in the quark-quark correlation function~\cite{Bashir:2012fs,Eichmann:2016yit}.
  The original motivation for diquarks has been the problem of `missing resonances' in the baryon spectrum which, despite predictions from the quark model, had not been observed in $\pi N$ scattering experiments.
  The addition of new states extracted from photo- and electroproduction experiments to the PDG~\cite{Olive:2016xmw} has ruled out early pointlike diquark models.
  On the other hand, sophisticated quark-diquark approaches derived from the three-body Faddeev equation produce a rich baryon spectrum that resembles the experimental situation~\cite{Eichmann:2016hgl}.
  It is therefore desirable to get a closer look at the properties of the baryons obtained in such a framework.

 \vspace{-3mm}

\section{Light baryons and their orbital angular momentum}
\label{sec:2}

   In Ref.~\cite{Eichmann:2016hgl} an effort was made to calculate the light baryon spectrum
   both from the covariant three-body Faddeev equation and its quark-diquark simplification,
   using the same underlying ingredients. The quarks and diquarks were not modeled but calculated
   from their Dyson-Schwinger and Bethe-Salpeter equations including their full covariant tensor decomposition.  
   The only input is~a~common rainbow-ladder interaction which models the $qq$ and $q\bar{q}$ kernel by an effective gluon exchange.

    \begin{figure*}[t]
    \centering
      \includegraphics[width=1\textwidth]{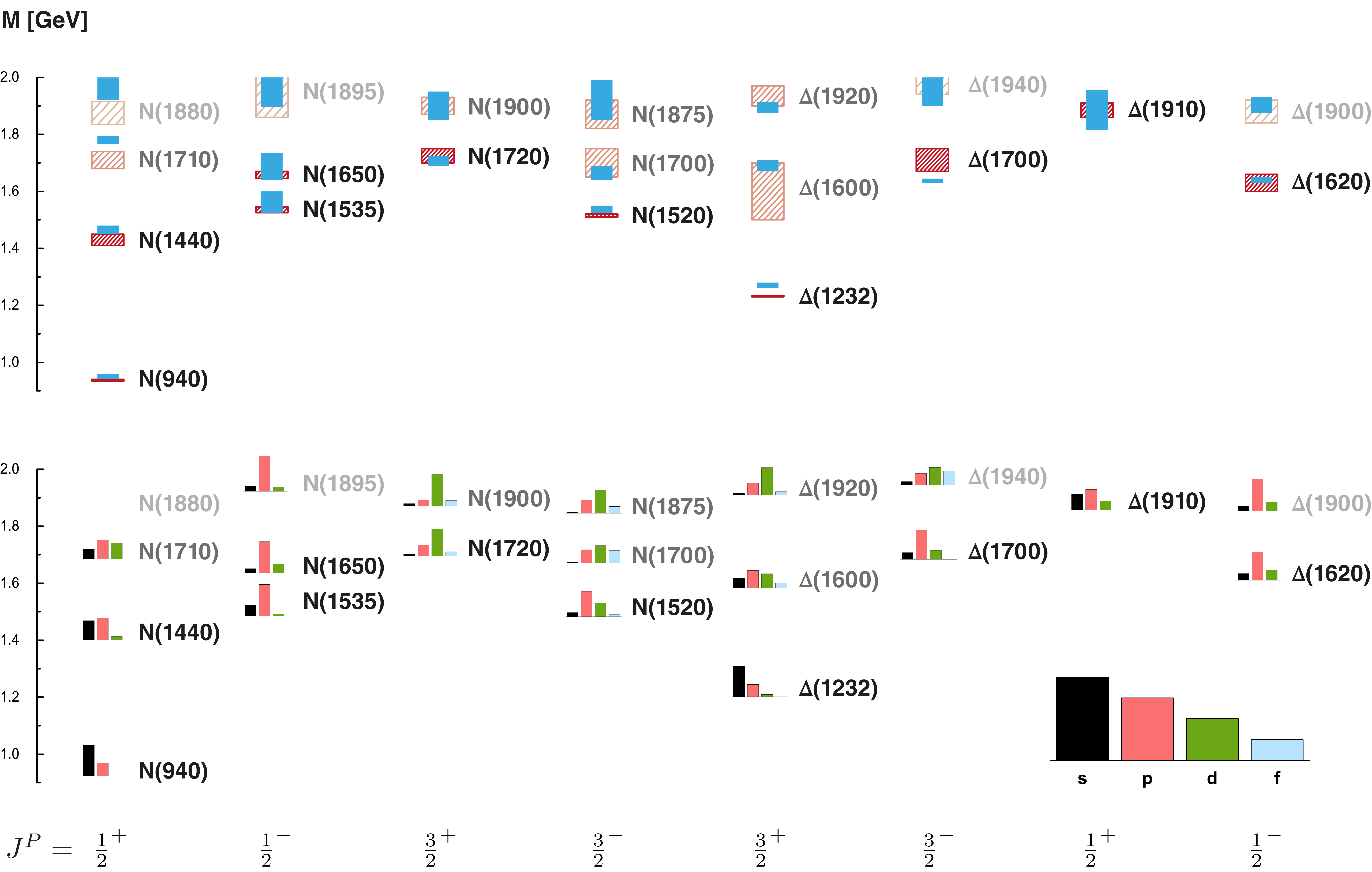}
    \caption{Light baryon spectrum in the quark-diquark approach with reduced pseudoscalar and vector diquarks.
             \textit{Top:} The calculated masses (solid boxes, blue) are compared to the PDG values (hatched boxes, red)~\cite{Olive:2016xmw}.
             \textit{Bottom:} Orbital angular momentum contributions (in $\%$) for the individual states; all bars sum up to $100\%$.}
    \label{fig:baryons}
    \end{figure*}

         As a result, the three-quark and quark-diquark results essentially agree with each other, at least for the low-lying states
         that are accessible in both cases. The rainbow-ladder truncation performs well in the $N(\nicefrac{1}{2}^+)$ and $\Delta(\nicefrac{3}{2}^+)$ channels but
         the remaining channels typically come out too low.
         In the quark-diquark system this behavior can be attributed to the underlying diquark properties: the `good' states are dominated by scalar and axialvector diquarks,
         whose meson counterparts (the pion and $\rho$~meson) are stable beyond rainbow-ladder. The pseudoscalar and vector diquarks, on the other hand, are bound too strongly
         and this affects the remaining channels.
         The strategy employed in Ref.~\cite{Eichmann:2016hgl} was to reduce the strength in these `bad' diquark channels by attaching a common constant to their Bethe-Salpeter kernels,
         which was adjusted by the $\rho$$-$$a_1$ splitting in the meson sector. As a result, the `bad' diquarks together with their (scalar and axialvector) meson counterparts
         acquire a higher mass and thereby resemble the outcome of beyond-rainbow-ladder calculations.
         The resulting baryon spectrum is shown in Fig.~\ref{fig:baryons} and agrees very well with the experimental spectrum,
         which suggests that a three-body calculation beyond rainbow-ladder should produce similar results.

         The resulting baryon Faddeev amplitudes carry a rich tensorial structure
         which can be organized into eigenstates of spin and orbital angular momentum in the baryon's rest frame~\cite{Eichmann:2016yit,Oettel:1998bk}.
         The orbital angular momentum content of a baryon can be quantified from the contributions to its Bethe-Salpeter normalization (see Eq.~\eqref{bs-norm} below).
         The resulting $s$, $p$, $d$ and $f$-wave components (corresponding to $L=0 \dots 3$) are shown in Fig.~\ref{fig:baryons}:
         \begin{itemize}
         \item the low-lying states in the $N(\frac{1}{2}^+)$, $\Delta(\frac{3}{2}^+)$ and $\Delta(\frac{1}{2}^+)$ channels carry sizeable $L=0$ components;
         \item the $N(\frac{1}{2}^-)$, $N(\frac{3}{2}^-)$, $\Delta(\frac{3}{2}^-)$, $\Delta(\frac{1}{2}^-)$ channels are typically dominated by $L=1$;
         \item the $N(\frac{3}{2}^+)$ channel is dominated by $L=2$.
         \end{itemize}
         In principle this is also expected from the nonrelativistic quark model, where the $SU(6) \times O(3)$ classification together with the relation $P=(-1)^L$
         entails that the nucleon and $\Delta$ belong to the $56$-plet with $L=0$ and positive parity, followed by the $70$-plet with $L=1$ and negative parity, and so on.
         Relativistically, however, these baryons carry all possible $L$ values allowed
         by $|L-S| \leq J \leq L+S$, without the restriction $P=(-1)^L$: only $J$ and $P$ are good quantum numbers but $L$ and $S$ are not,
         and therefore a relation between parity and orbital angular momentum cannot hold in general.
         Fig.~\ref{fig:baryons} shows that the nucleon and $\Delta(1232)$ have nonvanishing $p$-wave components, the $N(1535)$ has $s$ waves, and
         the $p$ waves appear to be even dominant for the Roper. 
         The subleading partial waves can have important consequences for form factors, for example in the $N\gamma\to\Delta$ transition~\cite{Eichmann:2011aa}.
         These results clearly demonstrate that light baryons are relativistic objects and Poincar\'e invariance is a necessary ingredient in their realistic description.

 \vspace{-3mm}

    \section{Complex conjugate eigenvalues?}

    A point noted in Ref.~\cite{Eichmann:2016hgl} is that some states in the calculated spectrum correspond to complex conjugate eigenvalues
    of the quark-diquark BSE. Their imaginary parts are small
    and thus they may simply be numerical artifacts,
    but from a conceptual point of view their appearance is not completely satisfactory.
    In fact, complex conjugate eigenvalues are a typical feature of BSEs for unequal-mass systems.
    They occur in scalar theories~\cite{Kaufmann:1969df,Seto:1992rc},
    where they can be associated with the interference of `normal' and `anomalous' states~\cite{Ahlig:1998qf},
    but for example also in heavy-light mesons~\cite{Rojas:2014aka,El-Bennich:2016qmb,Hilger:2016drj}.
    As we will see below, complex conjugate eigenvalues may even appear in equal-mass systems
    and therefore a closer inspection is desirable.

\begin{figure*}[t]
    \centering
      \includegraphics[width=1\textwidth]{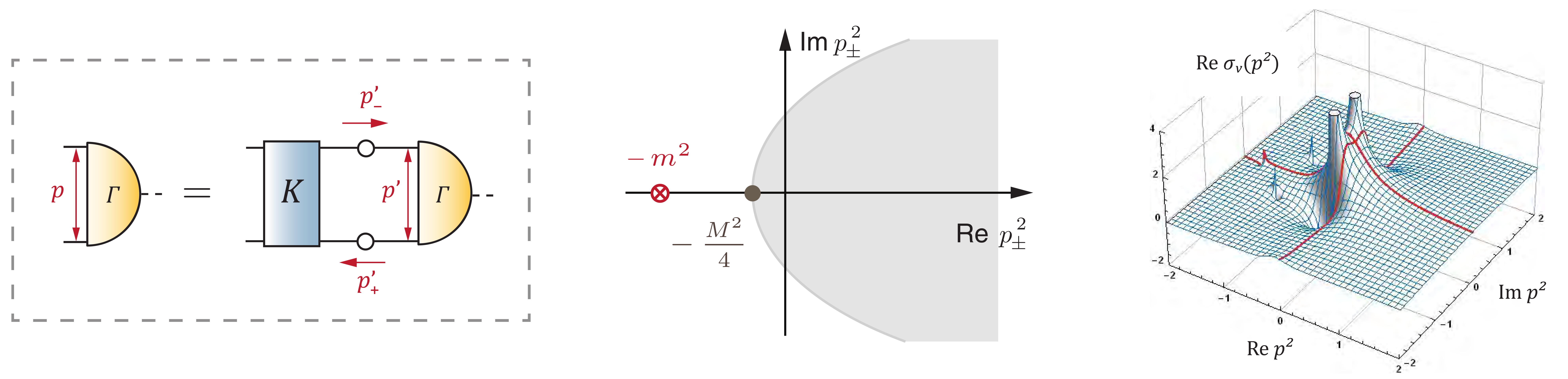}
    \caption{\textit{Left:} Generic form of a two-body Bethe-Salpeter equation. \textit{Center:} Parabolic domain of the propagator that is integrated over in the equation.
             \textit{Right:} Solution for the quark propagator in a rainbow-ladder truncation. Figures adapted from Ref.~\cite{Eichmann:2016yit}.}
    \label{fig:bse}
    \end{figure*}

    To provide a basis for the following discussion, let us quickly collect some basic formulas (more details can be found in~\cite{Eichmann:2016yit}).
    The generic form of a homogeneous BSE reads
    \begin{equation}\label{bse-generic}
        \mathbf{\Gamma}(p,P) =  \int \!\!\frac{d^4p'}{(2\pi)^4}\,K(p,p',P)\,G(p',P) \,\mathbf{\Gamma}(p',P)\,,
    \end{equation}
    where $\mathbf{\Gamma}(p,P)$ is the Bethe-Salpeter amplitude, $K$ is the kernel, and $G$ is the disconnected product of propagators as illustrated in Fig.~\ref{fig:bse}.
    $P$ is the total momentum ($P^2=-M^2$) and $p$, $p'$ are the external and internal relative momenta.
    The simplest example is a scalar theory with massive constituent particles of masses $m_{1,2}$ and $m=(m_1+m_2)/2$, which are bound by a scalar exchange particle with mass $\mu$ and coupling strength $c$:
    \begin{equation}\label{wc-1}
        K(p,p',P) = \frac{(4\pi m)^2\,c}{(p-p')^2 + \mu^2}\,, \qquad G(p,P) = \frac{1}{p_+^2+m_1^2}\,\frac{1}{p_-^2+m_2^2}\,, \qquad p_\pm = p + \frac{\varepsilon \pm 1}{2}\,P\,.
    \end{equation}
    For $\mu\to 0$ this becomes the well-known Wick-Cutkosky model~\cite{Wick:1954eu,Cutkosky:1954ru} which admits analytic solutions.
    $\varepsilon \in [-1,1]$ is an arbitrary momentum partitioning parameter and `ideal momentum partitioning' corresponds to $\varepsilon=(m_1-m_2)/(m_1+m_2)$.
    Here the amplitude $\mathbf{\Gamma}(p,P)$ is Lorentz invariant and therefore only depends on the variables $p^2$, $p\cdot P$ and $P^2=-M^2$.
    In practice it is convenient to factor out the angular dependencies in $z=\hat{p}\cdot\hat{P}$, for example with Chebyshev polynomials of the second kind,
    \begin{equation}\label{cheby}
        \mathbf{\Gamma}(p,P) = \sum_m \phi_m(p^2)\,U_m(z)\,, \qquad \frac{2}{\pi}\int dz\,\sqrt{1-z^2}\,U_m^\ast(z)\,U_n(z) = \delta_{mn}\,,
    \end{equation}
    which leads to an equation for the Chebyshev moments ($x=p^2$ and $x' = {p'}^2$):
    \begin{equation}
       \phi_m(x) = \sum_{nl} \int dx' \,K_{mn}(x,x')\,G_{nl}(x')\,\phi_l(x')\,.
    \end{equation}
    The resulting `kernel' and `propagator matrix' are given by (cf. Eq.~(3.44) in~\cite{Eichmann:2016yit})
    \begin{equation}\label{KG-WC}
    \begin{split}
        K_{mn}(x,x') &= \frac{2}{\pi}  \int dz \int dz'\,\sqrt{1-z^2}\,\sqrt{1-{z'}^2}\,\,U_m^\ast(z)\,U_n(z') \int \!\!\frac{dy}{(2\pi)^3} \,\,K(p,p',P) \,, \\
        G_{mn}(x) &= \frac{2}{\pi} \,\frac{x}{2}\int dz\,\sqrt{1-z^2}\,\,U_m^\ast(z)\,U_n(z)\,G(p,P)\,.
    \end{split}
    \end{equation}

    Upon discretizing also the radial variable $x$, the equation can be cast in the form
    \begin{equation} \label{bse-generic-2}
       K \,G\,\phi_i = \lambda_i\,\phi_i\,,
    \end{equation}
    which  generally holds for any homogeneous BSE.
    $K$ and $G$ are typically large matrices. For a two-body system
    their dimensions are of the order $\sim 10^3 \times 10^3$, depending on the number of grid points and invariant tensors (one for the scalar system, four for the pion, $\dots$)
    whereas for a three-body problem they become giant matrices of order $\gtrsim 10^6 \times 10^6$.
    The momentum dependencies in Eq.~\eqref{bse-generic} are encoded in the matrix dimensions of $K$ and $G$, except for the total momentum $P^2 = -M^2$ which remains an external variable.
    The equation is then solved for different values of $M$, which results in an eigenvalue spectrum $\lambda_i(M)$  of $KG$ and corresponding eigenvectors $\phi_i(M)$.
    At the intersections $\lambda_i=1$ one recovers the onshell Bethe-Salpeter amplitudes together with their masses, $P^2=-M_i^2$.

\begin{figure*}
    \centering
      \includegraphics[width=1\textwidth]{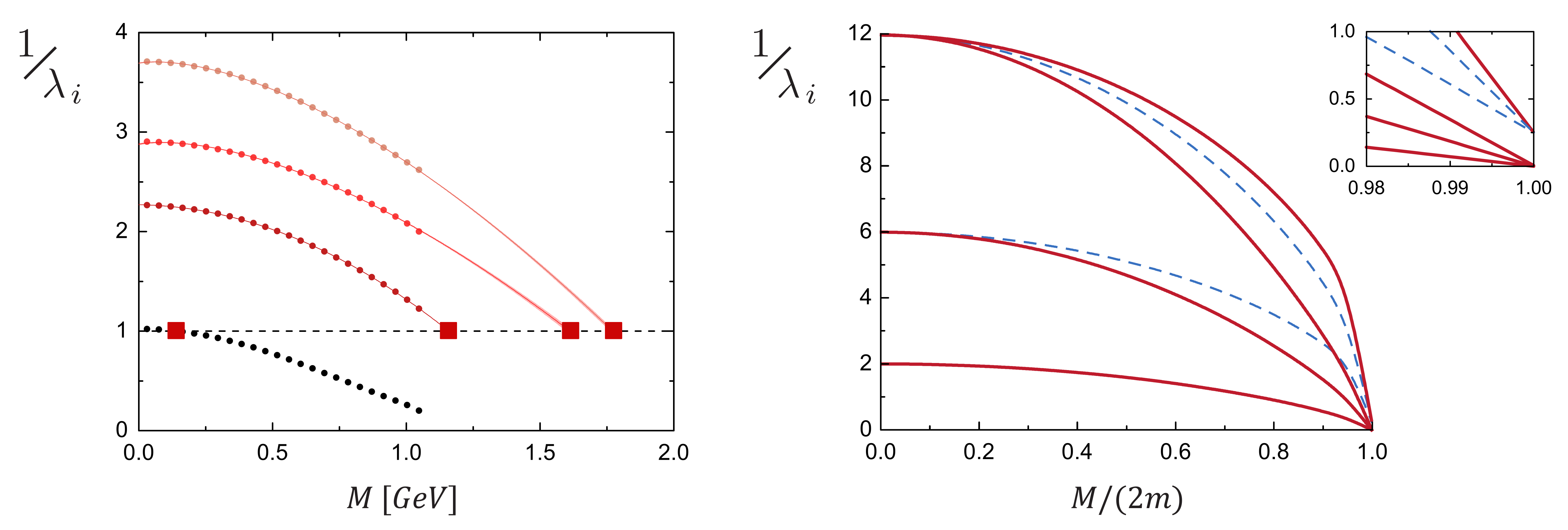}
    \caption{\textit{Left:} Eigenvalue spectrum for the pion channel with $J^{PC} = 0^{-+}$ obtained in rainbow-ladder. The boxes show the intersections $\lambda_i=1$ which
             correspond to the physical pion and its radial excitations. \textit{Right:} Eigenvalue spectrum for the Wick-Cutkosky model. Solid (red) curves are
             the solutions with positive `$C$ parity', dashed (blue) curves those with negative $C$ parity (see text).}
    \label{fig:ev}
    \end{figure*}

\begin{figure*}
    \centering
      \includegraphics[width=1\textwidth]{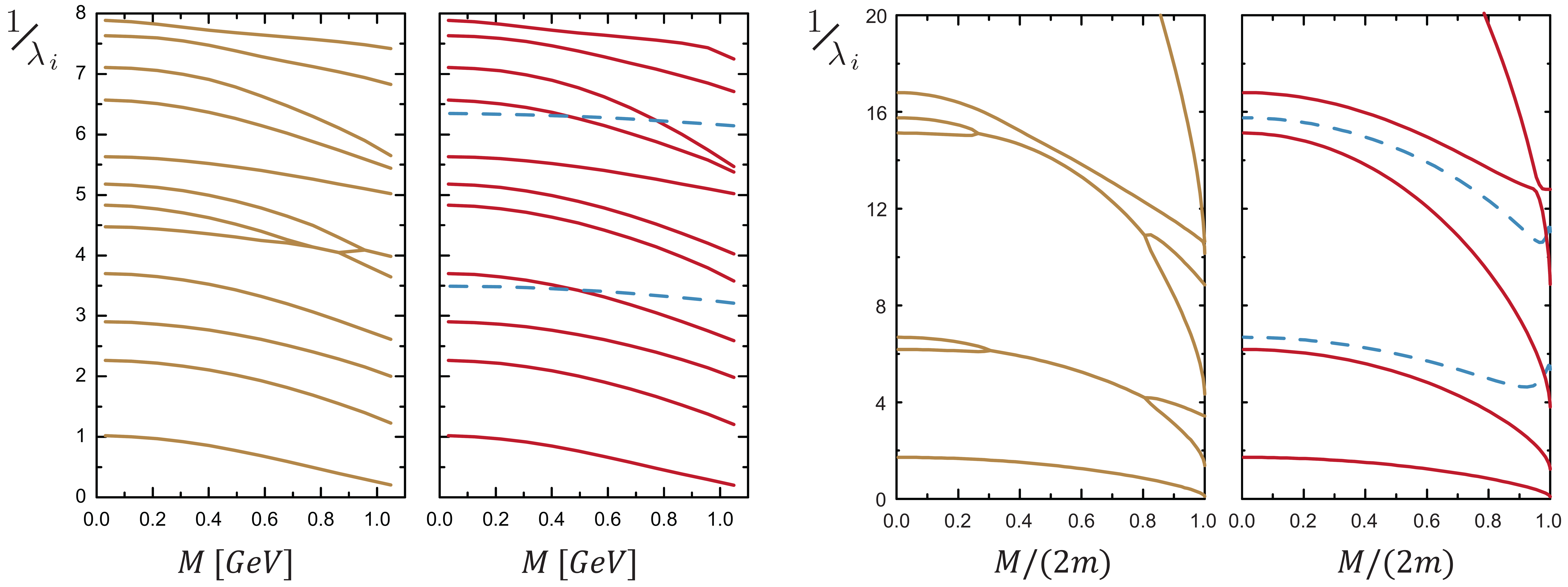}
    \caption{\textit{Left panels:} Eigenvalue spectrum for the pion before and after modifying the propagator matrix.
              The solid (red) lines are obtained by removing the negative eigenvalues of $G$ and the dashed (blue) lines by removing its positive eigenvalues.
             \textit{Right panels:} Analogous results for the Wick-Cutkosky model with unequal masses ($m_1/m_2=4$) and a nonvanishing exchange mass  ($\mu=m_2$).}
    \label{fig:wc}
    \end{figure*}

    An exemplary result is shown in the left panel of Fig.~\ref{fig:ev} for a pseudoscalar light quark-antiquark system with $J^{PC} = 0^{-+}$, calculated in rainbow-ladder.
    The corresponding BSE is given in Eq.~\eqref{bse-pion} below.
    The first eigenvalue $\lambda_0$ corresponds to the pion, the next one to its first radial excitation (experimentally, the $\pi(1300)$), and so on.
    The manifestation of spontaneous chiral symmetry breaking is visible in the plot: for massless quarks the eigenvalue for the ground state would start at $\lambda_0(M=0)=1$, so
    the pion becomes massless. In any case, it turns out that complex conjugate eigenvalues can occur even in the pion spectrum (see Fig.~\ref{fig:wc}): the first few eigenvalues are real,
    but in the higher-lying spectrum one encounters complex conjugate ones. How can one interpret them?

    In unequal-mass systems complex conjugate eigenvalues signal the loss of charge-conjugation invariance.
    This is related with the occurrence of anomalous states which appear for example in the Wick-Cutkosky model~\cite{Wick:1954eu,Cutkosky:1954ru,Nakanishi:1969ph}.
    The right panel in Fig.~\ref{fig:ev} shows the eigenvalue spectrum of the Wick-Cutkosky model for equal constituent masses $m_1=m_2=m$ and $\mu=0$.
    In this case there are no complex conjugate eigenvalues.
    In contrast to the case of the pion, the coupling $c$ in Eq.~\eqref{wc-1} is a free parameter and choosing a different value does not affect the propagators. We used $c=1$, but since a different choice
    only vertically rescales the eigenvalue spectrum one may as well replace the condition $1/\lambda_i=1$ for bound states by $1/\lambda_i = c$.

    In the Wick-Cutkosky model the constituent propagators have real mass poles at $p_\pm^2 = -m^2$.
    The propagators in the BSE are sampled within complex parabolas, cf.~Fig.~\ref{fig:bse}, from where one infers the condition $-m^2 < -M^2/4 \Rightarrow M < 2m$.
    This leads to a threshold, where the eigenvalues diverge at $M=2m$ and the inverse eigenvalues vanish --- except some that do not, which defines the anomalous states.
    Since they produce a vanishing binding energy ($M-2m=0$) for a nonvanishing coupling ($c=\nicefrac{1}{4}$), these states are unphysical.
    Anomalous states can be associated with excitations in relative time, which amounts
    to the distinction $\mathbf{\Gamma}(p,P) = \pm \mathbf{\Gamma}(-p,P)$ where the Bethe-Salpeter amplitudes
    are either even or odd in the angular variable $z$ and thus correspond to even or odd Chebyshev moments.

    The analogous symmetry for mesons is charge-conjugation invariance:
    \begin{equation}\label{cc}
        \conjg{\mathbf{\Gamma}}(p,P) = \pm \mathbf{\Gamma}(p,-P)\,, \qquad  \conjg{\mathbf{\Gamma}}(p,P) := C\,\mathbf{\Gamma}(-p,-P)^T\,C^T\,, \qquad C = \gamma^4 \gamma^2\,,
    \end{equation}
    which also leads to a distinction between even and odd Chebyshev moments.
    Equal-mass quarkonia are invariant under charge conjugation and their BSEs decouple for these two types of solutions, so they correspond to different physical particles
    with opposite $C$ parities: $J^{PC} = 0^{-+}$ for the pion and its radial excitations, and $J^{PC} = 0^{--}$ for members of the `exotic' channel.
    The latter do not have a non-relativistic limit due to the non-relativistic relations $P=(-1)^{L+1}$ and $\mathcal{C}=(-1)^{L+S}$. Relativistically
    these constraints no longer hold, for the same reasons as discussed in Sec.~\ref{sec:2}, and therefore exotic quantum numbers are in general not forbidden as $q\bar{q}$ states.
    The distinction between the pion and its exotic partner then amounts to nothing more than, for instance, the distinction between the $a_1$ ($1^{++}$) and the $b_1$ ($1^{+-}$) meson.

    In the scalar system charge-conjugation invariance can be easily verified: flipping the signs of $p$ and $p'$ in Eq.~\eqref{bse-generic} leads back to the same equation for $\mathbf{\Gamma}(-p,P)$
    because $K(-p,-p',P) = K(p,p',P)$ and, for equal constituent masses, $G(-p,P) = G(p,P)$. One can then
    form symmetric and antisymmetric combinations whose equations decouple.
    Without constraining the symmetry properties in advance, the eigenvalue spectrum is a superposition of both types of solutions.
    In the meson spectrum of Fig.~\ref{fig:ev} we already removed the $0^{--}$ eigenvalues, whereas in the Wick-Cutkosky spectrum we display both solutions:
    the solid (dashed) curves are those with positive (negative) `$C$ parity'. One can see that
    there is no 1:1 correspondence between the `wrong' $C$ parity and anomalous states.
    It has also been argued that anomalous states can indeed survive three-dimensional reductions
    and that the non-relativistic spectrum contains remnants of these states~\cite{Bijtebier:1997ir}.
    Hence, the remaining characteristic feature is that their eigenvalues remain finite at threshold.
    Ref.~\cite{Ahlig:1998qf} concluded that the anomalous states in the Wick-Cutkosky model only appear for values of the coupling
    that clash with the renormalization of the underlying quantum field theory.
    In Fig.~\ref{fig:ev} one can see that they only appear in the physical mass spectrum if $c \geq \nicefrac{1}{4}$;
    the consistent renormalization (which involves solving Dyson-Schwinger equations for the propagators)
    restricts $c$ to tiny values so they can never produce physical bound states and thus do not contribute poles to the S-matrix.
    In this sense anomalous states can be viewed as an artifact of the inconsistent combination of a ladder truncation with the usual tree-level propagators.

    What Fig.~\ref{fig:ev} also makes clear is that the definition of anomalous states in the above sense only applies to systems without confinement.
    The rainbow-ladder solution for the quark propagator produces complex conjugate poles (see Fig.~\ref{fig:bse}).
    Although this may not be the final word on the analytic structure of the quark propagator, it still poses a sufficient criterion for quark confinement due to positivity violations~\cite{Alkofer:2000wg}.
    Consequently, the meson eigenvalues in Fig.~\ref{fig:ev} do not diverge at $M=2m \sim 1$~GeV, where $m$ is defined from the parabola
    that is bounded by the first complex conjugate pole pair (in analogy to the center graph in Fig.~\ref{fig:bse}).
    Hence, there is no threshold and the mesons do not decay into free quarks,
    which also allows one to extrapolate their eigenvalues beyond the singularity limit.

    In summary it appears that neither of the existing definitions of anomalous states is truly applicable to QCD. QCD exhibits confinement and does not have quark thresholds, and
    it is a relativistic quantum field theory that may produce bound states without a nonrelativistic limit.
    Perhaps the only remaining characteristic of anomalous states is in terms of eigenvalues of $KG$  that never reach $\lambda_i=1$ and thus do not produce poles in the S-matrix.
    Whether the appearance of such eigenvalues in BSE solutions is truly a problem or not remains an open question;
    but even if so, it is not necessarily an inherent flaw of the theory because they may equally be truncation artifacts.
    Another possible criterion for unphysical states concerns the signs in spectral functions~\cite{Bhagwat:2007rj} which are determined by the slopes of the eigenvalues $\lambda_i$;
    see the comment after Eq.~\eqref{bs-norm-2} below.

    Let us now return to the question of complex eigenvalues.
    The typical situation in a BSE is that $K$ and $G$ are both Hermitian but $KG \neq (KG)^\dag$ because $K$ and $G$ do not commute.
    For unequal mass systems charge-conjugation invariance is lost, and the remnants of the positive- and negative $C$-parity solutions
    may start to interfere with each other. As discussed in~\cite{Ahlig:1998qf}, this is the underlying mechanism  that
    can lead to pairs of complex conjugate eigenvalues.
    On the other hand, complex eigenvalues may also occur in the excitation spectrum of equal-mass bound states
    (although not in the scalar system).
    Can one find a criterion that enforces real eigenvalues?

    Eq.~\eqref{bse-generic-2} is a generalized eigenvalue problem since it can be cast in the form
    \begin{equation}
       K\,[G\,\phi_i] = G^{-1} \lambda_i\,[G\,\phi_i]\,.
    \end{equation}
    If $G^{-1}$ is Hermitian and positive definite, one can employ a Cholesky decomposition $G^{-1} = LL^\dag$ to
    transform the equation into
    \begin{equation}
       C\,[L^\dag  G\,\phi_i] = \lambda_i\,[L^\dag  G\,\phi_i]\,, \qquad C = L^{-1} K\,(L^\dag)^{-1}\,,
    \end{equation}
    where $C=C^\dag$ if $K=K^\dag$. Since the equation has the same eigenvalues $\lambda_i$, they must indeed be real and the eigenvectors orthogonal:
    \begin{equation}\label{ev-orthogonality}
       \phi_i^\dag\,G^\dag  LL^\dag G\, \phi_j = \phi_i^\dag \,G\,\phi_j = \delta_{ij}\,.
    \end{equation}

    One should emphasize that the $\lambda_i$ and $\phi_i$ are defined for any $P^2=-M^2$, not only at the onshell points $P^2=-M_i^2$,
    and their orthogonality holds for each value of $M$ separately (but in general not for states with different $M$).
    From the spectral decomposition of $C$ we then further conclude:
    \begin{equation}
       K = \sum_i \lambda_i\,\phi_i\,\phi_i^\dag\,, \qquad
       K^{-1} = \sum_i \frac{1}{\lambda_i}\,\varphi_i\,\varphi_i^\dag\,, \qquad
       G = \sum_i \varphi_i\,\varphi_i^\dag\,, \qquad
       G^{-1} = \sum_i \phi_i\,\phi_i^\dag\,,
    \end{equation}
    where $\varphi_i = G\,\phi_i$ is the vector corresponding to the Bethe-Salpeter wave function.
    Moreover, if we project the full $n-$body Green function $\mathbf{G}$ and its connected, amputated part (the $\mathbf{T}$-matrix)  onto the Bethe-Salpeter eigenbasis for fixed $J^{PC}$ we obtain the relations
    \begin{equation}
       \begin{array}{rl}
          \mathbf{G} &= G + G\,K\,\mathbf{G} \\[1mm]
          \mathbf{T} &= K + K\,G\,\mathbf{T}
       \end{array}\qquad \Rightarrow \qquad
       \mathbf{G} = \sum_i \frac{\varphi_i\,\varphi_i^\dag}{1-\lambda_i}  \,, \qquad
       \mathbf{T} = \sum_i \lambda_i\,\frac{\phi_i\,\phi_i^\dag}{1-\lambda_i}
    \end{equation}
    which illustrate that each bound state ($\lambda_i = 1$ at $P^2=-M_i^2$) leads to a pole in $\mathbf{G}$ and $\mathbf{T}$.
    They also allow us to derive the proper \textit{normalization} from the residues at the pole positions:
    \begin{equation}
       \lambda_i = 1 + (P^2+M_i^2)\,\frac{d\lambda_i}{dP^2}\bigg|_{P^2=-M_i^2} + \dots \,,
    \end{equation}
    so that the onshell Bethe-Salpeter amplitudes and wave functions pick up normalization factors $g_i$:
    \begin{equation}\label{bs-norm-2}
       \mathbf{G} \; \stackrel{P^2\to -M_i^2}{\longlonglongrightarrow}  \; g_i^2 \,\frac{\varphi_i\,\varphi_i^\dag}{P^2+M_i^2} \,, \qquad
       g_i^2 = \left[ -\frac{d\lambda_i}{dP^2}\right]_{P^2=-M_i^2}^{-1}\,.
    \end{equation}
    Therefore, $g_i^2$ is always positive as long as $1/\lambda_i(M)$ approaches $1$ from above as in Fig.~\ref{fig:ev}  --- the opposite
    behavior would be another signature of an unphysical state.
    Eq.~\eqref{bs-norm-2} is equivalent to the standard Bethe-Salpeter normalization condition which can be written in the form~\cite{Nakanishi:1969ph}
    \begin{equation}\label{bs-norm}
       -g_i^2 \,\frac{d\lambda_i}{dP^2}\,\phi_i^\dag\,G\,\phi_i\,\Big|_{P^2=-M_i^2} \stackrel{!}{=} 1 \,.
    \end{equation}
    Finally, a similar decomposition holds for the vertex function $\psi$, defined by $G\,\psi = \mathbf{G}\,\psi_0$, which satisfies
    the inhomogeneous BSE $\psi = \psi_0 + K\,G\,\psi$:
    \begin{equation}
       \psi = \sum_i \frac{\alpha_i\,\phi_i}{1-\lambda_i} \; \stackrel{P^2\to -M_i^2}{\longlonglongrightarrow} \; g_i^2\,\frac{\alpha_i\,\phi_i}{P^2+M_i^2}\,, \qquad \alpha_i = \phi_i^\dag\,G\,\psi_0\,.
    \end{equation}

    The previous relations (real eigenvalues, orthogonal eigenvectors, etc.) hold as long as $K=K^\dag$, $G=G^\dag$, and $G$ is positive.
    In fact, the positivity of $G$ is only a sufficient but not a necessary criterion because it can be relaxed in several ways:
    we could have equally employed a Cholesky decomposition for $K$, which means that either $G$ \textit{or} $K$ should be positive;
    and if $G^{-1}$ had only negative eigenvalues, we could still write $G^{-1}=-LL^\dag$ to arrive at a Hermitian matrix $(-C)=(-C)^\dag$.
    In any case, the occurrence of complex conjugate eigenvalues of $KG$ signals that one or more of these criteria may be violated.
    In the following we will therefore review these assumptions in more detail.

     \textit{Symmetry:}
     The symmetry properties $K=K^T$, $G=G^T$ are not difficult to satisfy. The symmetry of $G$ is already ensured by the BSE itself;
          for example, in Eq.~\eqref{KG-WC} the condition $G_{mn}(x) = G_{nm}(x)$ is trivial because $z \in [-1,1]$ and the Chebyshev polynomials are real on that domain.
     The symmetry of $K_{mn}(x,x')=K_{nm}(x',x)$, on the other hand, leads to $K(p,p',P) = K(p',p,P)$ which can be implemented by a symmetrized kernel.
     Pictorially it refers to the mirror symmetry of $K$ in Fig.~\ref{fig:bse} around the vertical axis. The symmetry properties carry over
     to other types of BSEs, for example for $J=0$ mesons as fermion-antifermion systems:
          \begin{equation}\label{bse-pion}
             [\mathbf{\Gamma}(p,P)]_{\alpha\beta} =  \int \!\!\frac{d^4p'}{(2\pi)^4}\,[K(p,p',P)]_{\alpha\gamma,\beta\delta}\,[S(p'_+)\,\mathbf{\Gamma}(p',P)\,S(p'_-)]_{\gamma\delta}\,.
          \end{equation}
    In this case the equation can be transformed into a set of Lorentz-invariant integral equations by projecting the amplitude onto its Dirac basis elements
    (see Sec.~3.4 in Ref.~\cite{Eichmann:2016yit} for explicit examples):
    \begin{equation}\label{meson-decomposition}
       [\mathbf{\Gamma}(p,P)]_{\alpha\beta} = \sum_i f_i(p^2, z, P^2=-M^2)\,[\tau_i(p,P)]_{\alpha\beta}\,, \qquad \frac{1}{4}\,\text{Tr}\{\conjg{\tau_i}\,\tau_j\} = \delta_{ij}\,,
    \end{equation}
    where we used an orthonormal basis and charge conjugation was defined in Eq.~\eqref{cc}.
    The resulting propagator matrix is again symmetric,
    \begin{equation}
    \begin{split}
        G_{ij}(p,P) &= \frac{1}{4}\,\text{Tr}\{ \conjg{\tau}_i(p,P)\,S_A(p_+)\,\tau_j(p,P)\,S_B(p_-)\}
                     = \frac{1}{4}\,\text{Tr}\{ S_B^T(p_-)\,\tau_j^T(p,P)\,S_A^T(p_+)\,\conjg{\tau}^T_i(p,P)\} \\
                    &= \frac{1}{4}\,\text{Tr}\{ \conjg{\tau}_j(-p,-P)\,S_A(-p_+)\,\tau_i(-p,-P)\,S(-p_-)\}
                     = G_{ji}(-p,-P) = G_{ji}(p,P) \,,
    \end{split}
    \end{equation}
    and the same is true for its Chebyshev moments.
    In the last step we exploited the Lorentz invariance of $G_{ij}(p,P)$, which depends on $p^2$, $z$ and $P^2=-M^2$ only.
    Note that we only employed the \textit{definition} of charge conjugation but not charge-conjugation invariance itself, so
    the symmetry property also holds for heavy-light systems as highlighted by the subscripts $A$ and $B$ on the propagators.

    \textit{Reality:} that $K$ and $G$ are real is less obvious and in general also not the case.
    In the presence of thresholds the eigenvalues can become complex as the masses are shifted into the complex plane;
    for instance by an explicit decay mechanism in the kernel $K$, or potentially already
    in $G$ if a threshold in the propagators is crossed.
    In any case, the threshold locations are known in advance and below them the eigenvalues of $KG$ should be real.
    In the left panel of Fig.~\ref{fig:ev} one expects real eigenvalues for all values of $M$
    because a rainbow-ladder gluon exchange does not provide a decay mechanism for mesons and the quarks have complex conjugate poles,
    whereas for the Wick-Cutkosky model the eigenvalues should be real and positive below the threshold at $M=2m$.

    Still, even in the absence of thresholds $K$ and $G$ are not necessarily real. For example,
    depending on the $\tau_i$ in Eq.~\eqref{meson-decomposition} the dressing functions $f_i$ can be either real or imaginary;
    or in unequal-mass systems the propagator product $G(p,P)$ becomes complex.
    In these cases, however, one can recast Eq.~\eqref{bse-generic-2} in the form
    \begin{equation}\label{hh}
       (h \,K \,h)\,(h^\dag G \,h^\dag)\,(h\,\phi_i) = \lambda_i\,(h\,\phi_i)\,,
    \end{equation}
    where the vectors $h\,\phi_i$ are real and $h$ is a diagonal matrix whose entries are either $1$ or~$i$.
    As a consequence, $h \,K \,h$ and $h^\dag G \,h^\dag$ are real and symmetric matrices.
    The same reasoning applies for unequal-mass systems:
    with the Chebyshev expansion~\eqref{cheby}, odd Chebyshev moments are imaginary and therefore $K_{mn}(x)$ and $G_{mn}(x)$ are only symmetric and not real,
    but combinations such as $i^m G_{mn}(x)\,i^n$ are both real and symmetric.
    (It is customary to include a factor $i^m$ directly in the Chebyshev expansion, but this is still not enough because in that case $K_{mn}$ and $G_{mn}$ are real but not symmetric.)
    In general, $K$ and $G$ are not Hermitian but only \textit{pseudo}-Hermitian because they satisfy $K^\dag = h^2 K\,h^2$ and $G^\dag = h^2 G\,h^2$,
    where $h^2$ has diagonal entries $\pm 1$.
    In these cases, however, one can redefine $h \,K \,h \to K$, $h^\dag G \,h^\dag \to G$ and $h\,\phi_i \to \phi_i$ to arrive at real and symmetric matrices $K$ and $G$,
    which leads back to~Eq.~\eqref{bse-generic-2}.

    \textit{Positivity:}
    Neither of these alterations have changed the original equation~\eqref{bse-generic-2}.
    The leftmost plot in Fig.~\ref{fig:wc} shows again the pion spectrum, but now also including the eigenvalues for the higher-lying states.
    We have ensured in the calculation that $K$ and $G$ are strictly real and symmetric and yet we occasionally encounter complex conjugate eigenvalues which split into real branches.
    We are therefore left with the third option, namely that $G$ is not positive.
    To investigate this, we calculated the complete set of eigenvalues $\eta_i$ and eigenvectors $\xi_i$ of $G$
    and reconstructed the propagator matrix from its spectral decomposition, where we retained its positive eigenvalues only:
    \begin{equation}\label{spec-rec}
       G\,\xi_i = \eta_i\,\xi_i\,, \qquad \rightarrow \qquad G' = \sum_{\eta_i>0} \eta_i\,\xi_i\,\xi_i^\dag\,.
    \end{equation}
    We then insert the new propagator matrix into the original equation~\eqref{bse-generic-2},
    \begin{equation}
       K\,G'\,\phi_i' = \lambda_i'\,\phi_i'\,,
    \end{equation}
    and find that the new eigenvalues $\lambda_i'$ are indeed all real.

    It turns out that for equal-mass mesons removing the negative (positive) eigenvalues of $G$ also removes the negative (positive) $C$-parity states in the spectrum.
    This could have equally been achieved by restricting to even or odd Chebyshev moments; however,
    the procedure accomplishes more than that.
    Fig.~\ref{fig:wc} for the pion is already restricted to positive $C$-parity states.
    In the second panel we show the new eigenvalue spectrum obtained by removing either the negative eigenvalues of $G$ (solid, red)
    or its positive eigenvalues (dashed, blue).
    Removing the negative eigenvalues of $G$ does not change the low-lying spectrum, but it \textit{does} remove the
    complex eigenvalues in the higher-lying spectrum.
    Hence, all eigenvalues become real and the eigenvectors are automatically orthogonal in the sense of Eq.~\eqref{ev-orthogonality}.
    The `spurious' eigenvalues resemble the anomalous states in the Wick-Cutkosky model
    in the sense that they show much less curvature, which implies that they may not produce physical states.

    To test the effect on an unequal-mass system, we investigated the Wick-Cutkosky model with $m_1/m_2 = 4$ and a nonvanishing exchange mass $\mu = m_2$.
    The third panel in Fig.~\ref{fig:wc} displays the original eigenvalue spectrum, where the splitting of complex conjugate eigenvalue pairs into two real branches is clearly visible
    (see also Fig.~5 in Ref.~\cite{Ahlig:1998qf}). In the right panel we show the new eigenvalue spectrum.
    The previous complex conjugate pairs are now encompassed by two real eigenvalues, where the smaller one bends to zero
    as $M\to 2m$ and the larger one turns upwards and therefore does not generate a bound state.
    Interestingly, the procedure also leads to avoided level crossings (visible close to the threshold), which can switch the nature of `normal' and `anomalous' states.
    The `normal' eigenvalues display sharp curvatures very close to the threshold which are sensitive to the numerical resolution.
    Since this makes the spectral reconstruction~\eqref{spec-rec} numerically expensive
    we cannot yet resolve whether they truly vanish at the threshold, although it appears to be likely.
    We also note that in the unequal-mass system  the restriction of $G$ to its positive or negative spectrum does not restrict the resulting Bethe-Salpeter amplitudes to
    even or odd Chebyshev moments because they still contain both of them.

 \vspace{-3mm}

    \section{Summary}

    We discussed the light baryon spectrum in a quark-diquark approach where the quarks and diquarks are calculated from their Dyson-Schwinger and Bethe-Salpeter equations in QCD.
    The baryons' rest-frame orbital angular momentum contributions show
    clear traces of the non-relativistic quark-model, but
    due to Poincar\'e covariance their Faddeev amplitudes contain all possible values of $L$
    and such relativistic effects are already important for the nucleon, the $\Delta$, and the Roper resonance.

    We furthermore investigated the nature of complex conjugate eigenvalues that can appear in the excitation spectrum of Bethe-Salpeter equations.
    Our findings support the conjecture of Ref.~\cite{Ahlig:1998qf} that complex eigenvalues are generated by the interference of `normal' and `anomalous' solutions.
    Although it is not clear whether a rigorous criterion for anomalous states exists in QCD,
    one may associate them with eigenvalues that do not produce poles in the S-matrix.
    We find that a restriction of the propagator matrix to its positive (or negative) spectrum disentangles
    the complex conjugate eigenvalues of the Bethe-Salpeter equation, so they become real and exhibit the characteristics of normal (or anomalous) solutions.
    It will be interesting to apply this strategy to the spectra of heavy-light mesons and, in particular, to the light baryon spectrum.

    \bigskip

    \small
    \noindent
    \textbf{Acknowledgments:}
    I am grateful for interactions with C. S. Fischer, S.-X. Qin, H. Sanchis-Alepuz, and J.~Segovia.
    This work was funded by DFG Project No. FI 970/11-1 and by the DFG collaborative research center TR 16.

 \vspace{-3mm}


\bibliographystyle{spbasic}


\end{document}